# Wavelength dependence of photoinduced deformation in BiFeO$_3$

B. Kundys,[1,*] M. Viret,[2] C. Meny,[1] V. Da Costa,[1] D. Colson,[2] and B. Doudin[1]

[1]*Institut de Physique et de Chimie des Matériaux de Strasbourg (IPCMS), UMR 7504 CNRS-ULP, 67034 Strasbourg, France*
[2]*Service de Physique de l'EtatCondensé, DSM/IRAMIS/SPEC, CEA Saclay URA CNRS 2464, 91191 Gif-Sur-Yvette Cedex, France*

Optomechanical effects in polar solids result from the combination of two main processes, electric field-induced strain and photon-induced voltages. Whereas the former depends on the electrostrictive ability of the sample to convert electric energy into mechanical energy, the latter is caused by the capacity of photons with appropriate energy to generate charges and, therefore, can depend on wavelength. We report here on mechanical deformation of BiFeO$_3$ and its response time to discrete wavelengths of incident light ranging from 365 to 940 nm. The mechanical response of BiFeO$_3$ is found to have two maxima in near-UV and green spectral wavelength regions.

Cross-functional materials, where structure, charge, and magnetism are strongly interrelated, present high interest opportunities to realize new functionalities. In particular, photoelasticity in electrically polar materials is an extraordinary property with valuable remote light-controlled applications.[1–3] Due to the spontaneous polarization in polar dielectrics, light-generated charges distribute along the internal electric-field direction, thus changing the total electric field, and can cause sample deformation via the electrostrictive effect. Although some practical devices have been developed, the best response time with typical values of tens of seconds[4] must be improved. In that respect, our recent report on photoelastic effect in BiFeO$_3$ (BFO) with a fast response time may boost this research direction.[5] Moreover, the observed magnetic field dependence can offer additional functionality in future photoelastic-multiferroic devices where strain, magnetization, and polarization can potentially be changed simultaneously by light as well as applied magnetic and electric fields. The BFO is a well-documented magnetic and ferroelectric compound in which both properties have been investigated rather extensively.[6] Very interesting recent reports on optical properties in BFO include magnetochromism[7] and photovoltaic effects.[8,9] In this Brief Report, we investigate the wavelength dependence of the photoinduced strain in BFO single crystal for the purpose of finding an optimal cross-operational energy window of this intriguing property. Such measurements can also provide an independent indirect insight into the optical sensitivity of this extraordinary room temperature multiferroic compound.

A BFO single crystal with thickness of 90 $\mu$m and lateral dimensions in the mm scale, shown on Fig. 1, was used for this study. The sample was selected using polarized light with an optical microscope revealing its single ferroelectric domain state with its spontaneous ferroelectric polarization along the [111] direction in the pseudocubic lattice description.

Side 1 of the crystal is flat and reflective, allowing an easy verification of the sample ferroelectric state by polarized light in reflection mode, while side 2 exhibits a significant light absorption and indicates directions of crystalline axis with the longest dimension determined to be along [101].

Light emission diodes (LEDs) of several discrete 365, 455, 530, 660, 850, and 940 nm center wavelengths values with typical 30 nm spectral width, as well as a broadband white LED, were used as light source. We set these LEDs to equal power of 4.1 mW, measured with a Si diode power meter, using the power supply settings matching the provided manufacturer's specifications within 10%. This power was the minimal common value to the sources, chosen to minimize sample heating. The dimension change was measured along the [101] direction using a capacitance dilatometer similar to the one described in Ref. 10. Incoherent light was illuminating uniformly side 2 of the sample, along the [010] direction, with a distance between source and sample kept at 3.5 mm.

Upon irradiation of the sample with visible light of 365 nm with an irradiance of 326 W/m$^2$, a significant photoinduced striction was observed along the [101] direction (Fig. 2). An illumination time of 15 sec is chosen to compare the magnitude of the effect for different wavelengths, although for the 365-nm wavelength, the deformation does not reach full saturation within this time. Turning off the illumination results in a fast reversal of the elongation jump, followed by a very slow relaxation, noticeably not observed for longer excitation wavelengths (although also seen for the white light). The observed relaxation for 365-nm light has analogy with the slow capacitor discharge-like mechanism, which was also reported for other ferroelectric systems.[11] The sample typically needs at least approximately 4000 s to completely recover its initial size. The light-induced strain is tensile for all studied wavelengths under the same illumination power. As the wavelength increases, the observed deformation level shows a nonlinear behavior with a second maximum around 530 nm ($\sim$2.4 $\pm$ 0.12 eV) with further decreases for lower energies (Fig. 2, inset). The illumination with white light induces a midrange deformation value in agreement with the average deformation coming from the superposition of all wavelengths of the spectrum.

More intriguingly, the deformation response time is also found to be wavelength dependent (Fig. 3). Response time estimates were performed by fitting a linear approximation of the signal increase, as shown in Fig. 3 (inset), for all studied wavelengths. The shortest response time is observed for 365 nm and 530 nm, where the largest deformation level is found. Although the response time increases with increasing wavelength, it still remains much smaller than reported values for best-known PLZT photostrictive ferroelectric ceramics.[4] The response time due to illumination of white light is 1.7 s, which corresponds to an average of the response times for all discrete wavelengths measurements, ranging from 1.1 s to 2.7 s.

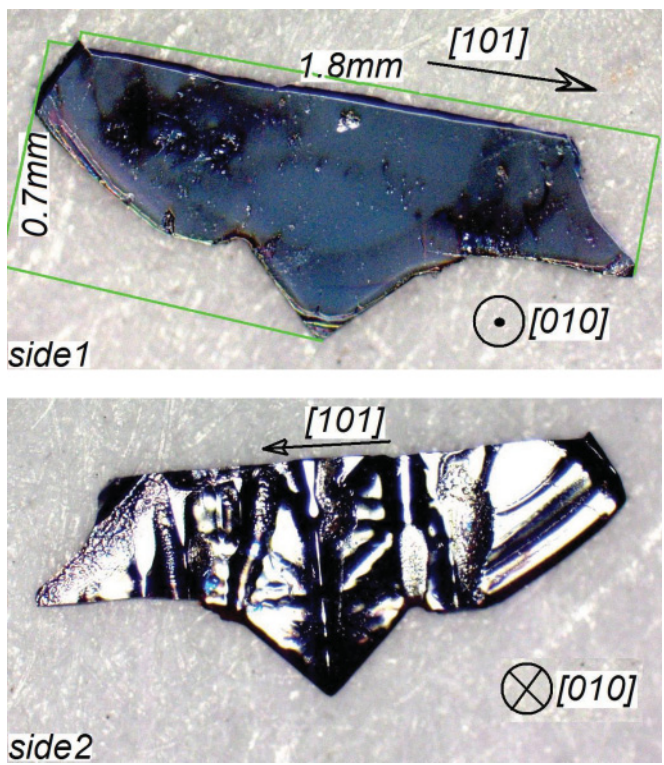

FIG. 1. (Color online) Optical microscope picture of a 90-$\mu$m-thick crystal of BFO. The longest edge is along [101].

The light with smaller light energy (530 nm) causes faster deformation than irradiance with larger energy (450 nm). This is in agreement with expected faster momentum transfer for energies closer to the bandgap of the material, where the photovoltaic effect can be large.[12] A faster response is also expected for larger photovoltaic currents.[13] The much larger deformation observed in the near UV region may be associated to the important role of the band-band transitions in BFO or impurity-defect absorption mechanisms. Indeed, the photocurrent in LiNbO$_3$:Fe was found to show maximum near the bandgap energy and then increased for light excitation energy exceeding the bandgap value.[14,15]

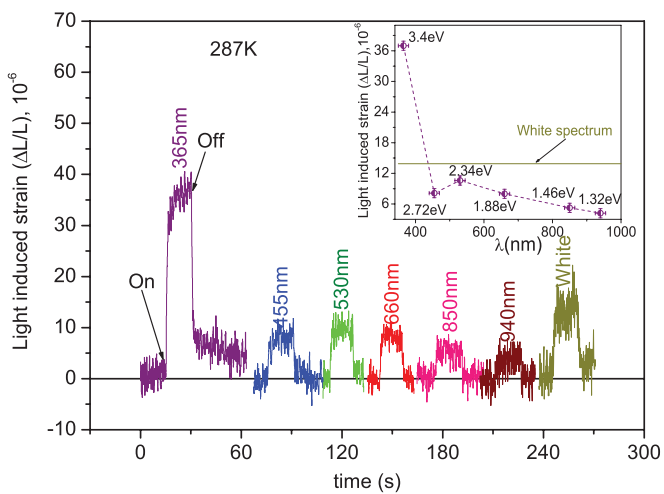

FIG. 2. (Color online) Photoinduced deformation of the BiFeO$_3$ crystal as a function of time. The inset shows the wavelength dependence of the average strain value taken during 15 s illumination time. The horizontal line represents the average strain level induced by white spectrum light (measured before all other wavelengths).

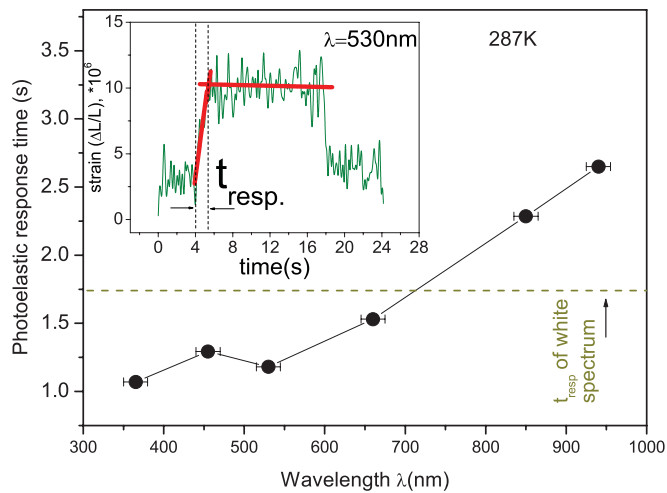

FIG. 3. (Color online) Response time of the photoelastic effect in the BiFeO$_3$ crystal as a function of wavelength. The horizontal dash line represents the BFO deformation response time induced by white spectrum light. The inset shows the response time determination approach using fits [red (dark gray) lines] for 530-nm wavelength.

The piezoelectric modulus can be deduced from the measured deformation $\tau$ as $\tau/E$, under the condition that the internal electric field $E$ of the sample is known (here the spontaneous electric field is likely to be along the polarization direction, i.e. either 71° or 109° with respect to the [100] axis). We took advantage of the photovoltaic properties of the sample to determine $E$. Semitransparent 20-nm-thick gold electrode contacts were deposited onto side 2 of our sample, while side 1 was covered with silver paste. The area of the gold electrode was 0.37 mm$^2$. The electric field $E$ produced by the light irradiation inside the sample can be extracted from $I(E)$ photocurrent measurements (Fig. 4). Because our measured deformation is transverse to the applied electric field, one has to use Poisson coefficient to determine the deformation longitudinal to the electric field. Although the Poisson ratio is not yet known for BiFeO$_3$ it can be expected to vary from 0.3 to 0.4.[16] Using a value of the electric field of 325 kV/m, light-induced deformation for green light ($9.5 \pm 1.8 \times 10^{-6}$) and the Poisson ratio taken from the aforementioned range, we estimate $d_{33} = (78 \pm 19)$ pm/V. The same order of magnitude was found in single crystal (16 pm/V)[17] and polycrystalline (44 pm/V)[18] BFO. The larger piezoelectric coefficient might also result from light-generated free carriers. This can be analyzed using Landau free energy expansion[19] adding the additional term $\sum_j n_j \varepsilon_j \Omega_j$, where $n_j$ is an average concentration of electron subsystem charges at the energy level $\varepsilon_j$ (conduction-band levels of traps and recombination) close to the bandgap energy and $\Omega_{ij}$ − stress tensor. The full expansion is of the form

$$F = \frac{1}{2}\alpha P^2 + \frac{1}{4}\beta P^4 + \frac{1}{6}\gamma P^6 - \frac{1}{2}\sum_i\sum_j k_{ij}\Omega_i\Omega_j$$
$$- P^2 \sum_j \xi_j \Omega_j + \sum_j n_j \varepsilon_j \Omega_j. \qquad (1)$$

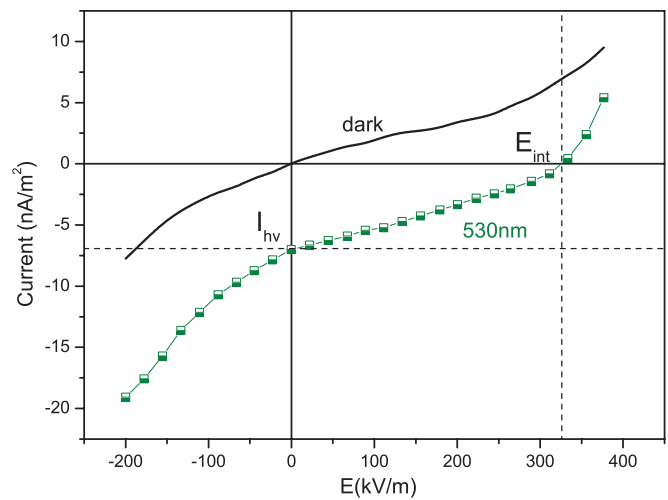

FIG. 4. (Color online) Current as a function of electric field in darkness and under 530 nm.

Here, $\alpha$, $\beta$, and $\gamma$ are known coefficients of the ferroelectric free-energy expansion, where $k_{ij}$ are the components of the elastic-stiffness tensor, and $\xi_j$ - are the components of electrostriction tensor. Taking the derivative $\partial F/\partial P$, which is the internal electric field, one can get

$$\frac{\partial F}{\partial P} = \alpha P + \beta P^3 + \gamma P^5 - 2P \sum_j \xi_j \Omega_j = E_j. \qquad (2)$$

The value of the strain $\tau_j$ can be defined as $\partial F/\partial \Omega_j$, where

$$\frac{\partial F}{\partial \Omega_j} = -\tau_j = -\frac{1}{2}\sum_i k_{ij}\Omega_i - P^2\xi_j + \sum_j n_j\varepsilon_j. \qquad (3)$$

Putting $\Omega_i = 0$ in Eq. (3), we find

$$-\tau_j \cong -P^2\xi_j + \sum_j n_j \varepsilon_j. \qquad (4)$$

This equation shows that the absolute strain does not depend on the sign of polarization and should be increased [the electrostriction being negative along the applied field direction [010] (Fig. 1)] in the presence of light-generated charges. Contraction in [010] (here $j$) direction corresponds to elongation along [101], which is indeed experimentally observed. Equation (4) provides a more fundamental qualitative description of photostriction in ferroelectrics. On the other hand, the piezoelectric coefficient determined as $|\tau_j|/E_j$ can be written from Eqs. (2) and (3) as follows:

$$d_{31} = \frac{-P^2\xi_j + \sum_j n_j\varepsilon_j}{\alpha P + \beta P^3 + \gamma P^5 - 2P\sum_j \xi_j\Omega_j}. \qquad (5)$$

This equation qualitatively illustrates how the piezoelectric modulus can be modified by light-induced charges. Since the absolute value of strain [the numerator of Eq. (5)] increases for negative electrostriction, the increase of piezoelectric coefficient can be expected. Moreover, photoinduced deformation can depend on free carriers-induced change in the electrostriction coefficient $\Delta\xi_j/\xi_j$ and lattice deformation described by the $\partial(\sum_j n_j\varepsilon_j)/\partial\Omega_j$ term (due to possible stress dependence of band energy levels[19]).

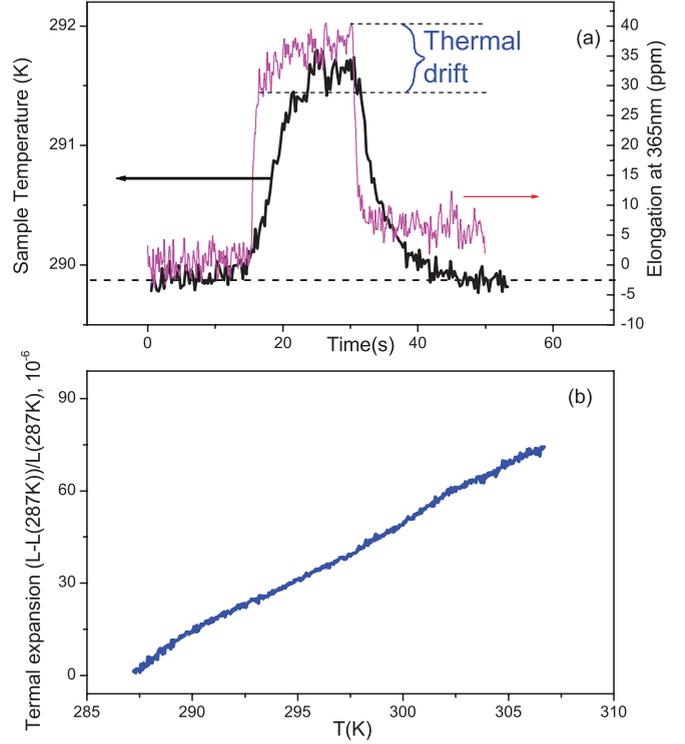

FIG. 5. (Color online) (a) Time dependence of the photostriction (right scale) and sample temperature change (left scale) due to light 365 nm. (b) Thermal expansion measured along [101] direction in BiFeO$_3$ crystal in darkness.

We also carefully checked that the light-induced temperature change did not create artifacts or bias in our conclusions. The sample was placed on the power-energy meter with temperature sensor (Thorlabs PM100D), which did not detect any measurable temperature change of the sample for light with wavelength $\geqslant$455 nm. We detected a temperature change not exceeding 0.4 K under 365-nm illumination using a Pt thermometer (attached to the sample), during an illumination time corresponding to the photoelastic response time of 1.1 s [Fig. 5(a)]. The measured thermal expansion with an external heater source [Fig. 5(b)] indicates that a sample heating by 9.5 K would correspond to the observed sample elongation under 365-nm excitation. This is more than one order of magnitude larger than the upper-bound estimate for light-induced heating. Further illumination reaches the maximum sample temperature change of 1.6 K. Taking into consideration the thermal expansion, this 1.6 K change corresponds to the thermal expansion along [101] of about 8 ppm [Fig. 5(b)]. Therefore, the thermal drift observed after saturation [Fig. 5(a)] (and absent for other wavelengths) can be attributed to the thermal expansion. This contribution is, however, not a principal one, and the observed wavelength-dependent times needed to deform the sample and to recover its initial shape (Fig. 1, Fig. 3 inset) are also difficult to reconcile with a model of dominant heating contribution to the observed effect.

In summary, the mechanical response of BiFeO$_3$ to incident light is reported in the 365- to 940-nm wavelength range. This reveals the most efficient operational energy window, found to have two maxima in the near-UV and the green lights. Our results can also bring complementary information on the electronic properties of BFO. In this respect it is interesting to note that reported values for the optical bandgap of BiFeO$_3$ range from 2.3 to 2.8 eV[20–26] at room temperature. The occurrence of two maxima of light-induced deformation near 2.34 and 3.4 eV suggests that other electronic transition energies can be predominant in BiFeO$_3$. Our findings of temporary (in the hour time range) irreversible changes that can occur under 365-nm illumination suggest that the region above UV should be measured carefully and better prior to near-UV measurements to ensure a complete virgin-state recovery (first exposure to light can show a bigger effect). Interestingly, there is also a debate regarding the direct[12,13] or indirect[27,28] nature of the bandgap in BiFeO$_3$. For direct bandgaps an electron can straightforwardly emit a photon. The present measurement concerns the opposite phenomenon: a photon generates an electron that is afterwards directed along the internal electric field of the crystal and causes the observed deformation. In that respect, the wavelength dependence of the response time, not observed previously in any system, provides an important complementary information. Although the magnitude of light-induced lattice deformation is rather small, it can reasonably be expected to be larger in samples with optimal thicknesses.[29] The reported order of magnitude increase in the piezoelectric properties in BFO doped with Tb,[30] Sm,[31] or La,[32] opens exciting opportunities to design materials with much larger photostriction (of the order of $10^{-4}$) providing bandgaps do not change drastically. Finally, the strongly nonlinear wavelength dependence and a large difference of the observed effects for the UV region suggest that this effect could possibly be used for UV-wavelength detection.

This work was partially supported by the ANR grant MELOIC. The technical help of the StNano cleanroom facility (H. Majjad) is gratefully acknowledged.